\documentclass[aps,pra,reprint,superscriptaddress]{revtex4-1}

\usepackage{siunitx}
\usepackage{graphicx}
\newcommand{\leftexp}[2]{{\protect\vphantom{#2}}^{#1}{#2}}

\begin{document}


\title{Imaging the evolution of an ultracold strontium Rydberg gas}


\author{P. McQuillen}
\email[]{patrick.mcquillen@rice.edu}
\affiliation{Rice University, Department of Physics and Astronomy and Rice Quantum Institute, Houston, Texas 77005 USA}

\author{X. Zhang}
\affiliation{Rice University, Department of Physics and Astronomy and Rice Quantum Institute, Houston, Texas 77005 USA}
\author{T. Strickler}
\affiliation{Rice University, Department of Physics and Astronomy and Rice Quantum Institute, Houston, Texas 77005 USA}
\author{F. B. Dunning}
\affiliation{Rice University, Department of Physics and Astronomy and Rice Quantum Institute, Houston, Texas 77005 USA}
\author{T. C. Killian}
\affiliation{Rice University, Department of Physics and Astronomy and Rice Quantum Institute, Houston, Texas 77005 USA}

\date{\today}

\begin{abstract}
Clouds of ultracold strontium $5\mathrm{s}48\mathrm{s} \ \leftexp{1}{\mathrm{S}}_0$ or $5\mathrm{s}47\mathrm{d} \ \leftexp{1}{\mathrm{D}}_2$ Rydberg atoms are created by two photon excitation of laser cooled $5\mathrm{s}^2 \ \leftexp{1}{\mathrm{S}}_0$ atoms. The spontaneous evolution of the cloud of low orbital angular momentum (low-$\ell$) Rydberg states towards an ultracold neutral plasma is observed by imaging resonant light scattered from  core ions, a technique that provides both spatial and temporal resolution. Evolution is observed to be faster for the S-states, which display isotropic attractive interactions, than for the D-states, which exhibit anisotropic, principally repulsive interactions. Immersion of the atoms in a dilute ultracold neutral plasma speeds up the evolution and allows the number of Rydberg atoms initially created to be determined.

\end{abstract}

\pacs{}

\maketitle

\section{Introduction}

By virtue of their strong long-range interactions and negligible thermal motion, ultracold Rydberg atoms provide many opportunities for studying interacting, many-body systems. Recent proposals have predicted that a rich variety of  phenomena should be observed, including  phase transitions to strongly correlated classical crystals \cite{bar08,ssl03,bdl07,pmb10,bsl11}, roton-maxon excitations \cite{ssl03,hnp10}, exotic spin and magnetic states \cite{les11}, three-dimensional solitons  \cite{mhs11}, and supersolids \cite{pro07,bon12,bpr12,pmb10,cjb10,hnp10}. Interactions can lead to large scale entanglement, collective Rabi oscillations, and  suppression of multi-atom excitations \cite{tfs04,sra04,hrb07}, which can be exploited for conditional logic operations and provide the basis for quantum information protocols \cite{jcz00,swm10}. Interactions can also give rise to strong nonlinear optical effects such as electromagnetically induced transparency \cite{pwa12}, and can lead to the formation of exotic long-range molecules involving Rydberg states \cite{bsc02,ost09,gds00,bbn09}. Attractive interactions can lead to collisions and Penning ionization \cite{rob05} and trigger a collisional cascade that can spontaneously transform a Rydberg gas into an ultracold neutral plasma (UNP) \cite{rtn00,kpp07,kpp07}.

Here, we study the spontaneous evolution of an ultracold ensemble of low orbital angular momentum (low-$\ell$) strontium Rydberg atoms  towards an ultracold plasma \cite{rtn00}, which is important because of the many new opportunities provided by alkaline-earth atoms for studies of Rydberg-atom physics.  The presence of a second valence electron leaves an optically active core after excitation of the Rydberg electron that can be used to further manipulate the Rydberg atom, such as through optical trapping  \cite{mmn11} or for optical imaging \cite{scg04}. The second electron also admits the possibility of creating doubly excited planetary atoms \cite{p77,trr00} or  autoionizing states \cite{cge78} that can serve as a probe of the evolution of the Rydberg electron \cite{mlj10}. Furthermore, the presence of singlet and triplet excited states allows the study of both attractive and repulsive Rydberg-Rydberg interactions within a single element \cite{vjp12}. Here we utilize the optically active core for \textit{in situ} spatial imaging of the evolution of a Rydberg gas, and demonstrate the sensitivity of this evolution to the sign and strength of the Rydberg-Rydberg interactions.

Most earlier experimental studies of the spontaneous evolution of an ultracold Rydberg gas into an ultracold plasma have utilized laser-cooled alkali or alkali-like atoms, including metastable xenon \cite{rbk98}, rubidium \cite{wgc04}, and cesium \cite{rtn00,wgc04}.
The principal diagnostic used in these experiments is the detection of charged particles that escape the plasma or are stripped from the atoms or plasma by an applied electric field.  Early work in this area \cite{vrg82} was motivated by the hope of observing an analog of the  Mott insulator-conductor transition \cite{mot74} in a dilute, amorphous system, but it was quickly realized that collisional processes dominate the evolution \cite{mck02,rha02,kon02prl,kon02,rha03}.
Recent experiments exploring the evolution of cold Rydberg molecules seeded in a supersonic jet  towards a plasma have extended measurements to extremely high densities where new collision processes such as associative ionization can become important \cite{mrk08}.

Recent experimental studies of ultracold strontium Rydberg atoms include spectroscopy \cite{mlc11} and the use of autoionization to detect changes in Rydberg-electron orbital angular momentum induced by collisions with free electrons \cite{mlj10}. Calculations
of long-range interactions \cite{vjp12} based on the known quantum defects \cite{czr99,bls82,awe79,rbo78,bsc83} predict
isotropic attractive  van der Waals interactions for $5\mathrm{s}n\mathrm{s} \ \leftexp{1}{\mathrm{S}}_0$ states, with a $C_6$ coefficient of comparable magnitude to that for rubidium $n\mathrm{s} \ \leftexp{2}{\mathrm{S}}_{1/2}$ states. In contrast, the strontium $5\mathrm{s}n\mathrm{d} \ \leftexp{1}{\mathrm{D}}_2$ states exhibit interactions that range from slightly attractive to repulsive with a mean strength about one-half that for the $5\mathrm{s}n\mathrm{s} \ \leftexp{1}{\mathrm{S}}_0$ state but of opposite sign \cite{vjp12}. The present experimental results support these calculations.


\section{Experimental Methods}
\subsection{Rydberg and Plasma Creation}
\begin{figure*}
\includegraphics[width=6in]{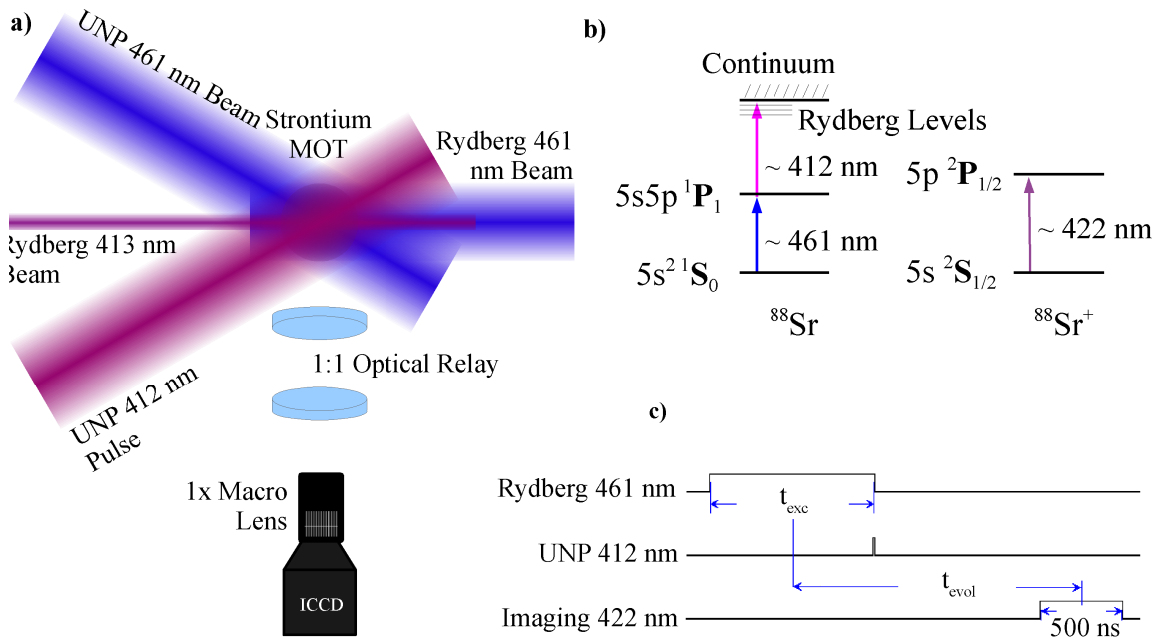}
\caption{a) Experimental schematic showing the Rydberg excitation beams and the UNP ionization beams used for creation of a seed plasma. For simplicity,  the Zeeman, cooling, and fluorescence-imaging beams are not shown. The latter are nearly collinear with the Rydberg excitation beams. b) Pertinent energy levels for the strontium atom and ion. c) Timing diagram. Rydberg excitation, for times $t_{exc}$ of \SIrange{1}{20}{\micro\second}, is followed by a \SI{10}{\nano\second} ionization pulse (if exciting an UNP). A variable evolution time ($t_{evol}$) of \SIrange{0.5}{10}{\micro\second} elapses before the \SI{500}{\nano\second} imaging pulse is applied. \label{fig:RydbergSetup}}
\end{figure*}

The present experiments are conducted in a magneto-optical trap (MOT) that is loaded using a Zeeman-slowed beam of $\mathrm{\leftexp{88}{Sr}}$ atoms and operated on the principal  $\leftexp{1}{\mathrm{S}}_0 - \leftexp{1}{\mathrm{P}_1}$ transition at \SI{461}{\nano\metre}.  Approximately $10^9$ atoms are captured, and the density varies as $n=n_0\mathrm{exp}(-r^2/2\sigma^2)$, with characteristic radius $\sigma =$ \SI{1}{\milli\metre} and peak density $n_0 \sim $ \SI{6e10}{\per\centi\metre\cubed}. The atom temperature is $\sim$ \SI{7}{\milli\kelvin}.

After loading, the trapping beams and MOT field coils are turned off and Rydberg atoms are created through two-photon excitation (Fig.\ \ref{fig:RydbergSetup}), using a combination of 461\,nm radiation red-detuned \SI{430}{\mega\hertz}  from the intermediate $^1\mathrm{P}_1$ level and \SI{413}{\nano\metre} radiation tuned to excite either the $5\mathrm{s}48\mathrm{s} \ \leftexp{1}{\mathrm{S}}_0$ or $5\mathrm{s}47\mathrm{d} \ \leftexp{1}{\mathrm{D}}_2$  states.
The \SI{461}{\nano\metre} light is generated by frequency doubling a Ti:sapphire laser that is stabilized via saturated absorption spectroscopy.
 The \SI{413}{\nano\metre} light is provided by a frequency-doubled diode laser, whose frequency is tuned and stabilized with the aid of an optical transfer cavity locked to a stabilized HeNe laser \cite{lsd91}. The two concentric, counterpropagating beams have the same (linear) polarization and $e^{-2}$ diameters of $\sim$ \SI{3.5}{\milli\metre} and $\sim$ \SI{2}{\milli\metre} and peak intensities of $\sim$ \SI{1.6}{\watt\per\cm\squared} and $\sim$ \SI{5.1}{\watt\per\cm\squared}, respectively. The \SI{413}{\nano\metre} light remains on for the entire experiment, and the time of excitation $t_{exc}$ is set by pulsing the \SI{461}{\nano\metre} beam with an acousto-optic modulator (AOM). The resulting cigar-shaped cloud of Rydberg atoms has peak densities as high as \SI{4e8}{\per\centi\metre\cubed}.


For some experiments, an ultracold neutral plasma is created immediately following Rydberg atom excitation through direct two-photon ionization (Fig.\ \ref{fig:RydbergSetup}) of a controllable fraction of the (ground state) atoms in the MOT. In the present work an ionization fraction (IF) of $\sim$ \SI{0.02}{\percent} is employed. The \SI{461}{\nano\metre} radiation is blue detuned by \SI{20}{\mega\hertz} and pulsed using an AOM, while the \SI{412}{\nano\metre} radiation that promotes electrons to just above the ionization continuum is generated by a \SI{10}{\nano\second} pulsed dye laser. Both laser beams are larger than the atom cloud and the density profile follows that of the MOT, with  electron and ion densities, $n_e\sim n_i \sim $ \SI{1.3e7}{\per\cm\cubed}. This is a much lower density than typically used in strontium UNP experiments \cite{kpp07}, but it is well above the threshold for electron trapping and plasma formation \cite{kkb99}. The energy, $E_e$, of the photoelectrons sets the electron temperature, $T_e$, given by $T_e \equiv \frac{2}{3} E_e/k_B =$ \SI{105}{\kelvin}, a temperature consistent with the observed expansion dynamics \cite{lgs07}. The heavy ions inherit their initial temperature from the laser cooled atoms, however, they heat within a few hundred nanoseconds to about \SI{2}{\kelvin} due to disorder-induced heating \cite{mur01,scg04,csl04}.

\subsection{Imaging Diagnostics}
\label{sec:ID}

Typically the evolution of a population of Rydberg atoms is monitored using state selective field ionization (SFI) \cite{gal94}. However, while SFI provides a measure of state selectivity, it provides no spatial resolution
except in special circumstances such as when used with Rydberg atoms in large magnetic fields \cite{cgp05} or  in conjunction with highly magnifying charged particle optics \cite{ssr11}.
Here we demonstrate a new diagnostic for studying gases of ultracold Rydberg atoms based on imaging resonant light scattered off the  core ions that can provide spatially and temporally resolved \textit{in situ} measurements.  The present apparatus provides temporal resolution of $\sim$ \SI{10}{\nano\second} and spatial resolution of $\sim$ \SI{20}{\micro\metre} \cite{mck11}, although the latter could be improved to potentially approach the diffraction limit in an optimally designed apparatus. Although the LIF technique does not provide the capacity for state discrimination afforded by SFI, it can, as demonstrated here, be used to follow collisional dynamics and the evolution of Rydberg populations. Alkaline-earth atoms are particularly attractive for optical imaging as they possess core ions with principal transitions in the visible or near-UV. (Similar techniques are employed for imaging UNPs \cite{scg04,cgk08}).

For the strontium UNPs and Rydberg atoms used here, laser-induced fluorescence (LIF) imaging and spectroscopy is performed on the $5\mathrm{s} \ \leftexp{2}{\mathrm{S}}_{1/2}$ - $5\mathrm{p} \ \leftexp{2}{\mathrm{P}}_{1/2}$ core-ion transition at \SI{422}{\nano\metre}. The LIF excitation beam is linearly polarized and is formed into an ellipsoidal cross section, with $e^{-2}$ intensity diameters along the major and minor axes of $\sim$ \SI{7}{\milli\metre} and $\sim$ \SI{1}{\milli\metre}. It interacts with a narrow  ``sheet" of atoms or plasma perpendicular to the imaging axis and parallel to the Rydberg excitation beams. The probe laser is on for $\sim$ \SI{0.5}{\micro\second} and its peak intensity, $\sim$ \SI{3.5}{\milli\watt\per\centi\metre\squared}, is three times the saturation intensity for a bare ion. Scattered photons are imaged onto an intensified CCD which provides spatially and temporally resolved two-dimensional images. Varying the frequency of the LIF laser yields the full, spatially resolved, excitation spectrum. While measurements of the Doppler broadening can be used to obtain velocity distributions, we focus here on measurements of density profiles.

A full excitation spectrum is needed to quantitatively measure the density of visible core ions because broadening (or shifting) of the spectrum as a result of Doppler effects and/or interactions between  the core ion and Rydberg electron can change the peak scattering intensity \cite{cgk08,mlj10}. Corrections must also be applied to account for the finite size of the LIF beam; by considering the sizes of the LIF beam and the core-ion density distribution \cite{cgk08} the full density distribution of the visible core ions can be derived from the images (with an assumption about the size along line of site). The total number of core ions imaged can then be determined by integrating the measured density over space.

When the LIF excitation spectrum is constant or only slowly varying, as is the case for most of the present work, a spatially resolved signal proportional to density can be obtained from the fluorescence at a single frequency on peak resonance. This greatly reduces the amount of data that must be taken, and is the approach typically used here. We calibrate the measured on resonance LIF intensity against the true density by recording the full spectrum for a subset of experimental conditions and determining the appropriate calibration factor \footnote{Absolute calibration of the LIF signal is obtained using absorption imaging \cite{scg04,cgk08}.}.

\section{Results and Discussion}
\subsection{Low-$\ell$ Rydberg Atoms and the Effect of an UNP Seed}
\begin{figure}
\includegraphics[width=3.375in]{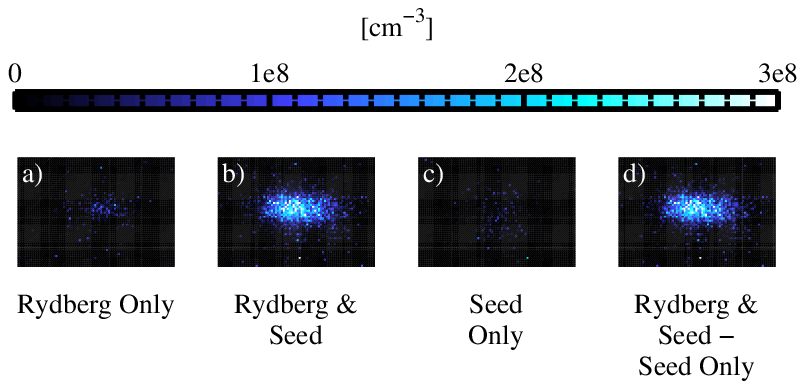}
\caption{Laser-induced fluorescence imaging of UNP and Rydberg atom clouds using the
$5\mathrm{s} \ \leftexp{2}{\mathrm{S}}_{1/2}$ - $5\mathrm{p} \ \leftexp{2}{\mathrm{P}}_{1/2}$ core-ion transition at \SI{422}{\nano\metre}. (a) Image after Rydberg excitation  to the $5\mathrm{s}48\mathrm{s} \ \leftexp{1}{\mathrm{S}}_0$ state for $t_{exc} \approx$ \SI{3}{\micro\second}, which yields $\sim$ \num{8e5} Rydberg atoms. Notice that the fluorescence signal is very small. (b) Image after exciting the same Rydberg population as in (a) but with superposition of a seed UNP containing $\sim$ \num{2e5} ions/electrons. Note the increased fluorescence from the cigar-shaped region of Rydberg excitation. (c) Image of an identical seed UNP to that in (b), but with no Rydberg excitation. (d) Signal due to Rydberg excitation obtained by subtracting the signal due to ion cores in the seed UNP. \label{fig:Images}}
\end{figure}

 Optical images provide a sensitive spatially resolved probe of collision dynamics, i.e., of whether the parent Rydberg atoms remain in their initial low-$\ell$ states. The series of images shown in Fig.\ \ref{fig:Images} help demonstrate the capabilities of the present diagnostic and its sensitivity to collisional effects.

For atoms excited to low-$\ell$ Rydberg states, such as the $5\mathrm{s}48\mathrm{s} \ \leftexp{1}{\mathrm{S}}_0$ and $5\mathrm{s}47\mathrm{d}\ \leftexp{1}{\mathrm{D}}_2$  states, core excitation leads to autoionization of the Rydberg atom through electron-electron scattering on a time scale on the order of the classical electron orbital period, $\sim$\SI{19}{\pico\second} at $n \sim 50$ \cite{cge78,gal94}. This results in extreme broadening of the core transition compared to a bare ion. The oscillator strength, determined predominantly by the ground and excited state wavefunctions of the core electron, is affected very little by the Rydberg electron, so extreme broadening  greatly reduces  the peak transition rate. Extrapolating references \cite{xzm86} and \cite{xzm87}, we estimate autoionization spectral widths $(\gamma_{AI})$ of \SI{10}{\giga\hertz} and \SI{30}{\giga\hertz} for the $5\mathrm{s}48\mathrm{s} \ \leftexp{1}{\mathrm{S}}_0$ and $5\mathrm{s}47\mathrm{d}\ \leftexp{1}{\mathrm{D}}_2$  states, respectively.  For our \SI{422}{\nano\metre} laser intensity, core excitation for S and D Rydberg states is thus very weak and unlikely during our 500\,ns LIF pulse. Our probe is not sensitive to atoms in these states, which is consistent with the data presented in Fig.\ \ref{fig:Images}(a) showing  very little fluorescence.


As $\ell$ increases the autoionization rate drops rapidly due to the diminishing overlap between the Rydberg electron and core-ion wavefunctions. By $\ell\approx 8$  \cite{gal94,mlj10} the autoionization rate drops below the core radiative decay rate, allowing fluorescence to begin being observed.  With further increases in $\ell$, the autoionization rate becomes negligible and excitation of the core ion leads directly to fluorescence.  In essence, the overlap between the excited Rydberg electron and core ion becomes so small that the core ion behaves as an independent, isolated particle.

Thus Rydberg atoms can be optically imaged by transitioning them to high-$\ell$ states.  This can be accomplished in a controlled manner by adding a small seed UNP, which introduces free electrons that quickly promote Rydberg electrons to higher angular momentum states through quasi-elastic $\ell$-mixing electron-Rydberg collisions \cite{dfw01}. The cross section for this process when the velocity of the free electron is close to the classical orbital velocity of the Rydberg electron is very large \cite{dfw01}. Figure \ref{fig:Images}(b) shows how the presence of a relatively small ``seed" UNP leads to a dramatic increase in the LIF signal. A Rydberg-UNP interaction time greater than \SI{0.5}{\micro\second} was allowed before imaging, which, based on the work in Ref. \cite{wgc04}, results in strong $\ell$-mixing.

Given the high degeneracy associated with high-$\ell$ states for the values of $n$ of interest here, and the fact that $\ell$-mixing collisions lead to a statistical population of final states \cite{gal94}, few low-$\ell$ states remain and it is reasonable to assume that in the presence of an UNP all Rydberg atoms become visible through LIF after $\ell$-mixing collisions have occurred, allowing their number and density to be determined.  At later times collisional processes beyond simple $\ell$-changing become important including  $n$-changing and collisional ionization.  However, once $\ell$-changing has occurred, these processes do not change the visibility of the ion cores.  Indeed, in no case was a decrease in the fluorescence signal seen at late times.

In the presence of an UNP, the LIF signal contains contributions from both the Rydberg ion cores and the ions formed in creation of the seed UNP.  The latter contribution, however, can be determined from measurements with no Rydberg atoms present (Figure \ref{fig:Images}(c)). The larger spatial extent and relatively low density of the UNP (\SI{1e7}{\per\centi\metre\cubed}) as compared to the Rydberg cloud \SIrange{1e8}{4e8}{\per\centi\metre\cubed} facilitates accurate and complete subtraction of its contribution, as demonstrated in Fig.\ \ref{fig:Images}(d).

\subsection{Rydberg Atom Excitation Rates}
\begin{figure}
\includegraphics[width=3.375in]{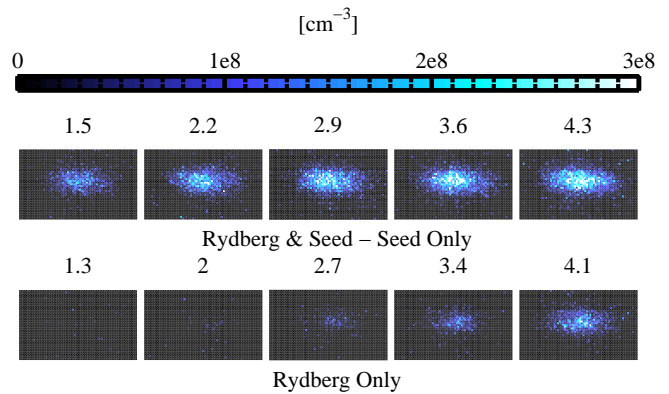}
\caption{Dependence of the LIF signal from parent  $5\mathrm{s}48\mathrm{s} \ \leftexp{1}{\mathrm{S}}_0$ Rydberg atoms on the excitation time $t_{exc}$ (shown above each image in \si{\micro\second}) with (top) and without (bottom) a seed UNP present. (Contributions to the LIF signals from the UNP are subtracted.) \label{fig:Density}}
\end{figure}

\begin{figure}
\includegraphics[width=3.375in]{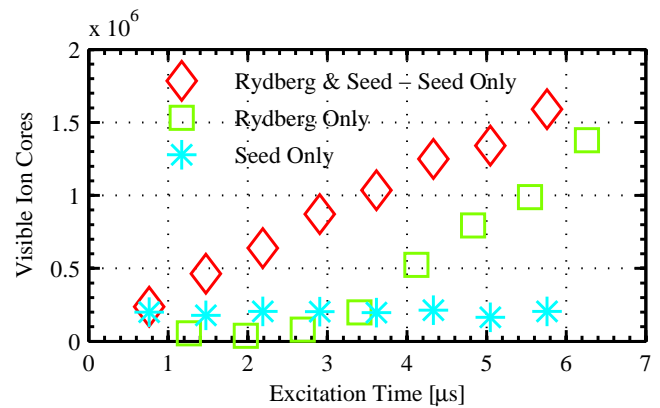}
\caption{Number of visible ion cores versus Rydberg excitation time for parent $5\mathrm{s}48\mathrm{s} \ \leftexp{1}{\mathrm{S}}_0$ states. Data recorded both with and without a seed UNP present are included together with results obtained when only the seed UNP is created. Qualitatively similar results are seen for the $5\mathrm{s}47\mathrm{d} \ \leftexp{1}{\mathrm{D}}_2$ state. \label{fig:Excitation}}
\end{figure}

The rapid $\ell$-mixing induced by the presence of an UNP allows direct measurement of the number of Rydberg atoms initially excited.  This is illustrated in the upper panels in Fig. \ref{fig:Density} which show the results of measurements in which the excitation pulse length, $t_{exc}$, was varied.  Immediately following excitation a seed UNP was created and, after a further delay of \SI{0.5}{\micro\second}, the $\ell$-mixed Rydberg atom cloud was imaged by LIF.  The resulting Rydberg density images obtained after the subtraction of the contribution to the LIF signal from the UNP are presented in Fig. \ref{fig:Density}. Figure \ref{fig:Excitation} shows a quantitative analysis of this study in which the observed density is integrated to give the total number of visible  core ions, i.e., the number of Rydberg atoms excited.  As expected, this number increases linearly with excitation time.  Figure \ref{fig:Excitation} also includes the results of measurements of the number of ions created in the UNP,  which is time independent.

As illustrated in the lower panels of Fig. \ref{fig:Density}, a very different time dependence of the LIF signal is observed when no UNP is present.  Very few visible ion cores are seen for excitation times less than $\sim$ \SI{3}{\micro\second} even though, as shown in Fig. \ref{fig:Excitation}, close to \num{1e6} Rydberg atoms have been created.  This lack of visibility indicates that, on this time scale, the Rydberg atoms remain in low-$\ell$ states.  An increasing number of core ions, however, become visible for longer excitation times. For the longest excitation times the number of visible core ions is seen to approach that observed when an UNP is present.

\subsection{Spontaneous Evolution of a Rydberg Gas into an UNP}

\begin{figure}
\includegraphics[width=3.375in]{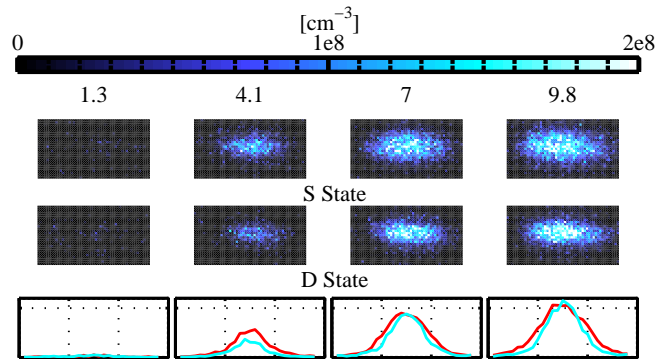}
\caption{LIF images showing the spontaneous evolution of an ultracold gas of $\leftexp{1}{\mathrm{S}}_0$ (top) and $\leftexp{1}{\mathrm{D}}_2$ (middle) Rydberg atoms. The evolution time is indicated above each image in \si{\micro\second}. The initial numbers and densities for both states are identical, \num{8e5} and \SI{2.2e8}{\per\centi\metre\cubed}, respectively. Notice how the S-state population evolves more quickly in both space and time. The bottom panels show one-dimensional plots of the transverse density integrated along the horizontal direction. The spatial development of the S-state (red)  leads that of the D-state (blue). \label{fig:SvsDDensity}}
\end{figure}

The free evolution of an ultracold gas of parent $5\mathrm{s}48\mathrm{s} \ \leftexp{1}{\mathrm{S}}_0$ atoms is shown in the top panels in Fig.\ \ref{fig:SvsDDensity}.  In these measurements an initial excitation time, $t_{exc} \approx $ \SI{2.5}{\micro\second}, was selected as this yields a large Rydberg atom population ($\sim$ \num{7e5} atoms) with little core ion visibility (see Fig. \ref{fig:Density}) indicating that little $\ell$-changing has occurred.  This collection of atoms was then allowed to evolve freely for some time, $t_{wait}$, following which the core ions were imaged.  To roughly account for dynamics during excitation and imaging, the evolution times shown in Fig.\ \ref{fig:SvsDDensity} are taken as  $t_{evol}= t_{exc}/2 + t_{wait} + t_{ccd}/2$ where $t_{ccd} = $ \SI{.5}{\micro\second} is the exposure time of the image.  As evident from Figs.\ \ref{fig:SvsDDensity}  and \ref{fig:SvsD} the number of visible core ions increases dramatically with increasing evolution time, pointing to $\ell$-changing and the spontaneous evolution of the parent ultracold Rydberg gas towards an UNP. Imaging the  core ions provides a spatial and temporal probe of the early stages of this dynamics.

The spontaneous evolution of a Rydberg gas towards an UNP has attracted significant experimental \cite{klk01,lnr04,wgc04} and theoretical attention \cite{pvs08,kon02,ppr03,rha03} because of the complex dynamics involved and its importance for experiments involving clouds of interacting Rydberg atoms \cite{swm10}. A key step in this process is the initial generation of free electrons. Various mechanisms have been suggested as contributing to this, including interactions between Rydberg atoms,  ionization by blackbody radiation, and collisions with hot background gas \cite{cpi10}. For Rydberg atoms with attractive interactions, the consensus is that interactions dominate, as has been demonstrated experimentally \cite{wgc04,lnr04,ltg05} and described theoretically \cite{rob05}. Nearby Rydberg atoms accelerate towards each other, and when their separation approaches the classical electron orbital radius, Penning ionization occurs. This produces a more deeply bound Rydberg atom and a free electron with kinetic energy on the order of the initial binding energy. Initially, these electrons pass quickly out of the atomic cloud, leaving their core ions behind, until a Coulomb potential forms that traps subsequently formed electrons \cite{kkb99}. For the present geometry it is estimated that this requires creation of $\sim$\num{2e3} free ions. Once this has occurred the $\ell$-mixing rate  increases rapidly \cite{vct05}, which quickly results in formation of a fully $\ell$-mixed population.

The time at which electron trapping and $\ell$-mixing begin can be estimated by considering the time required for sufficient numbers of nearest neighbor Rydberg atoms to collide and produce the required number of free ions.  This can be accomplished by adapting calculations similar to those found in ref. \cite{rob05} and using the $C_6$ coefficients given in ref. \cite{vjp12}, along with the Erlang distribution of nearest neighbor separations. In this manner we estimate this time for $5\mathrm{s}48\mathrm{s} \ \leftexp{1}{\mathrm{S}}_0 - \leftexp{1}{\mathrm{S}}_0$   attractive interactions to be $\sim$\SI{3}{\micro\second} for the present Rydberg densities of  $\sim$ \SI{2e8}{\per\centi\metre\cubed}. This is consistent with the time of marked increase in LIF signal (Fig. \ref{fig:SvsD}).

Note in Fig.\ \ref{fig:SvsDDensity} that LIF is initially confined to a central region that is smaller than the overall size of the Rydberg atom cloud.  This can be attributed to the higher atom density in this region which leads to larger collision rates and formation of a deeper Coulomb well to trap the Penning electrons.  As time increases the volume over which electrons are trapped grows steadily. In consequence, as seen in Fig. \ref{fig:SvsD}, the visible ion signal increases for several microseconds before reaching its final saturated value at which point all the initial core ions become visible.

Following $\ell$-changing, further collisions can lead to $n$-changing and to ionization, destroying the Rydberg atoms and creating an UNP.  The present diagnostic provides valuable information on the spatial and temporal evolution of the Rydberg gas towards high-$\ell$ states, which is an important step in the formation of an UNP, but it is insensitive to subsequent $n$-changing and collisional ionization.

\subsection{Rydberg Gas Dynamics with Attractive and Repulsive Interactions.}

To explore how sensitive the evolution of a Rydberg gas is to  the Rydberg-Rydberg interactions, data were also recorded using $5\mathrm{s}47\mathrm{d} \ \leftexp{1}{\mathrm{D}}_2$ states.  These states have anisotropic interactions that range from slightly attractive to repulsive \cite{vjp12}. To allow direct comparison to measurements undertaken using  $\leftexp{1}{\mathrm{S}}_0$ states, the laser intensities were adjusted to produce clouds of $\leftexp{1}{\mathrm{D}}_2$ atoms having the same size and shape and the same number of atoms, $\sim$ \num{8e5}.  To eliminate possible effects associated with Zeeman splitting induced by the decaying MOT magnetic field, the field-insensitive $m_j=0$ $\leftexp{1}{\mathrm{D}}_2$ state was excited.
	
\begin{figure}
\includegraphics[width=3.375in]{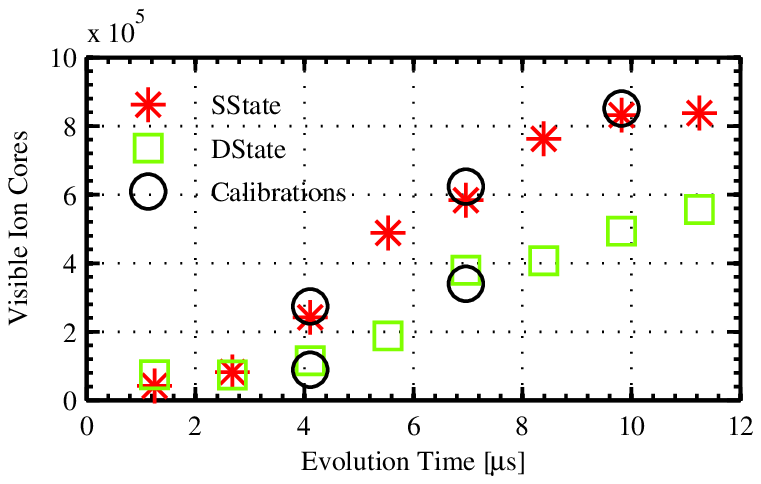}
\caption{Evolution of the number of visible core ions for $5\mathrm{s}48\mathrm{s} \ \leftexp{1}{\mathrm{S}}_0$ and $5\mathrm{s}47\mathrm{d} \ \leftexp{1}{\mathrm{D}}_2$ Rydberg atoms.  The circles represent number calibrations performed by scanning the imaging laser through resonance (see section \ref{sec:ID}). At late times the number of visible ion cores seen $\sim$ \num{8e5} agrees reasonably well with the number of parent Rydberg atoms initially excited, $\sim$ \num{7e5}, as determined using a seed UNP. \label{fig:SvsD}}
\end{figure}

As evident from Fig.\ \ref{fig:SvsDDensity}, while the general characteristics of the evolution of the $\leftexp{1}{\mathrm{D}}_2$ state cloud are similar to those for the  $\leftexp{1}{\mathrm{S}}_0$ state cloud, i.e., visibility starts in the densest regions and propagates in space and time to encompass the entire cloud, the time scale for this evolution is noticeably slower as evidenced by the smaller spatial extent and lower LIF intensities seen at any selected evolution time.  This is quantified in Fig.\ \ref{fig:SvsD} which shows the growth in the number of visible ion cores.  It is significantly slower for $\leftexp{1}{\mathrm{D}}_2$ than for $\leftexp{1}{\mathrm{S}}_0$ states consistent with weaker interactions and slower Penning ionization for  $\leftexp{1}{\mathrm{D}}_2$.

Rydberg atoms with repulsive interactions are known to Penning ionize and produce plasma  \cite{arw07,arg07,cpi10}, but at a much slower rate than for attractive potentials  \cite{arg07}. However, because this growth is slow, other processes must be considered that might contribute to the growth in the number of visible ion cores after $\leftexp{1}{\mathrm{D}}_2$ excitations, such as population redistribution to states with attractive potentials by blackbody radiation (BBR) and BBR-induced photoionization.  Calculations for hydrogen, helium, and the alkali metals \cite{btr07,gno10,ogn11} suggest that while the rates for excitation/deexcitation of Rydberg states by BBR depend on the particular atomic species, the rate for states with $n\sim 50$ should be less than $\sim$\SI{1e4}{\per\second}.  Whereas such rates might allow up to $10 \%$ of the parent $\leftexp{1}{\mathrm{D}}_2$ Rydberg states to undergo a transition to states of different $\ell$ on the time scale ($\sim$\SI{10}{\micro\second}) of the present measurements, this is insufficient to significantly increase the overall transition rate.  (Furthermore, because of the dipole selection rules, i.e. $\Delta \ell = \pm 1$, several sequential transitions are required to populate Rydberg states of sufficiently high $\ell$ that their ion cores become visible allowing this process to be neglected.) 

The calculated rates for BBR-induced photoionization again vary from atom to atom, but at $n \sim 50$ typically center around \SI{3e2}{\per\second}.  This implies that, given an initial population of  $\sim$\num{8e5} Rydberg atoms, times of $\sim$\SI{8}{\micro\second} would be required to create the $\sim$\num{2e3} atoms necessary to begin trapping electrons and initiate rapid $\ell$ mixing.  This seems insufficient to be the dominant contributor under our conditions, for which the onset of $\ell$ mixing begins at approximately \SI{3}{\micro\second}. It is not negligible however, and it might explain, at least in part, why the difference between the data sets recorded using (attractive) $\leftexp{1}{\mathrm{S}}_0$ and (principally-repulsive) $\leftexp{1}{\mathrm{D}}_2$ states is not larger.

\section{Conclusions}

We have demonstrated core-ion imaging, a new technique for studying the dynamics of ultracold Rydberg gases.  It can provide spatial as well as temporal information on the collision dynamics governing the evolution of an ultracold gas of interacting Rydberg atoms, in particular on the $\ell$-changing interactions that occur during the early stages of the evolution towards an UNP.  Furthermore, with the addition of an UNP seed, the total number and density of Rydberg excitations can be measured. Observations of the evolution of the Rydberg atom population demonstrate  sensitivity to the strength and sign of the interactions between neighboring Rydberg atoms and are consistent with recent predictions that interactions between $5\mathrm{s}48\mathrm{s} \ \leftexp{1}{\mathrm{S}}_0$ states are attractive whereas those between $5\mathrm{s}47\mathrm{d} \ \leftexp{1}{\mathrm{D}}_2$ states are largely repulsive.
The present technique promises new insights into the properties of ensembles of ultracold Rydberg atoms and could be particularly useful, for example, for imaging spatial correlations, such as have recently been seen in rubidium with an atomic microscope \cite{ssr11}.


\begin{acknowledgments}
We gratefully acknowledge financial support from the National Science Foundation (PHY-0964819), the Department of Energy Partnership in Basic Plasma Science and Engineering (PHY-1102516), the Air Force Office of Scientific Research (FA9550-12-1-0267), and the Robert A. Welch Foundation under grant No. C-0734.
\end{acknowledgments}

%

\end{document}